\documentclass[conference]{IEEEtran}
\ifCLASSINFOpdf
\else
\fi
\usepackage{algorithm}
\usepackage{algorithm,algpseudocode}
\usepackage{graphicx}
\usepackage{array}
\usepackage{float}
\usepackage{lipsum}
\usepackage{authblk}
\usepackage{mathtools}
\usepackage{authblk}
\usepackage{hyperref}

\usepackage{subcaption}

\begin{document}
%
\title{A Long Short-Term Memory Recurrent Neural Network Framework for Network Traffic Matrix Prediction}
%
%
%

\author{Abdelhadi Azzouni\thanks{abdelhadi.azzouni@lip6.fr}}

\author{Guy Pujolle\thanks{guy.pujolle@lip6.fr}}

\affil{LIP6 / UPMC; Paris, France  \{abdelhadi.azzouni,guy.pujolle\}@lip6.fr}

\maketitle

\begin{abstract}

Network Traffic Matrix (TM) prediction is defined as the problem of estimating 
future network traffic from the previous and achieved network
traffic data. It is widely used in network planning, resource management and network security.
Long Short-Term Memory (LSTM) is a specific recurrent neural network (RNN) architecture  
that is well-suited to learn from experience to classify, process and predict time series 
with time lags of unknown size. LSTMs have been shown to model temporal sequences and their 
long-range dependencies more accurately than conventional RNNs.
In this paper, we propose a LSTM RNN framework for 
predicting 
Traffic Matrix (TM) in large networks. 
By validating our framework on real-world data from G\'EANT network, 
we show that our LSTM models converge quickly and give state of the
art TM prediction performance for relatively small sized models.


\end{abstract}


 {\bf { \it keywords - }}
Traffic Matrix, Prediction, Neural Networks, Long Short-Term Mermory

%
\IEEEpeerreviewmaketitle

\section{Introduction}

Most of the decisions that network operators make
depend on how the traffic flows in their network.
However, although it is very important to accurately estimate traffic parameters,
current routers and network devices do not provide the possibility for real-time 
monitoring, hence network operators cannot react effectively to 
the traffic changes. To cope with this problem, prediction techniques have been 
applied to predict network parameters and therefore be able 
to react to network changes in near real-time. 

The predictability of network traffic parameters is mainly determined by 
their statistical characteristics and the fact that they present a strong correlation between
chronologically ordered values. Network
traffic is characterized by: self-similarity, multiscalarity, 
long-range dependence and a highly nonlinear nature (insufficiently modeled by Poisson and Gaussian models) \cite{selfsimilar}.

A network TM presents the traffic volume between all pairs of origin and
destination (OD) nodes of the network at a certain time t. 
The nodes in a traffic matrix can be Points-of-Presence (PoPs), 
routers or links.

Having an accurate and timely network TM is essential for most network 
operation/management tasks such as traffic accounting, short-time traffic scheduling or re-routing, 
network design, long-term capacity planning, and network anomaly detection.  
For example, to detect DDoS attacks in their early stage, it
is necessary to be able to detect high-volume traffic clusters in
near real-time. Another example is, upon congestion occurrence in the network, 
traditional routing protocols cannot react immediately to 
adjust traffic distribution, resulting
in high delay, packet loss and jitter. 
Thanks to the early warning, a proactive prediction-based approach will be faster, in terms
of congestion identification and elimination, than reactive
methods which detect congestion through measurements, only
after it has significantly influenced the network operation.

Several methods have been proposed in the literature for
network traffic forecasting. These can be classified into two
categories: linear prediction and nonlinear prediction. The most
widely used traditional linear prediction methods are: a)
the ARMA/ARIMA model \cite{forecasting}, \cite{studypredict}, \cite{mpegvid}  and b) the
Holt–Winters algorithm \cite{forecasting}. The most common nonlinear
forecasting methods involve neural networks (NN) \cite{forecasting}, \cite{mpeg2},
\cite{mpeg3}. The experimental results from \cite{evaluation}
show that nonlinear traffic prediction based on NNs 
outperforms linear forecasting models (e.g. ARMA, ARAR, HW)
which cannot meet the accuracy requirements. 
Choosing a specific forecasting technique is based on a 
compromise between the complexity of the solution, characteristics
of the data and the desired prediction accuracy.
\cite{evaluation} suggests if we take into
account both precision and complexity, the best results are
obtained by the Feed Forward NN predictor with multiresolution learning
approach.

 

%

Unlike feed forward neural networks (FFNN), 
Recurrent Neural Network
(RNNs) have cyclic connections over time. 
The activations from each time step are stored in the internal
state of the network to provide a temporal memory.
This capability makes RNNs better
suited for sequence modeling tasks such as time series prediction and
sequence labeling tasks. 

Long Short-Term Memory (LSTM) is a RNN architecture that was designed by 
Hochreiter and Schmidhuber \cite{lstmoriginalpaper} to  address the  
vanishing and exploding gradient problems of conventional RNNs.
RNNs and LSTMs have been successfully used for handwriting recognition \cite{handwritinglstm}, 
language modeling, phonetic labeling of acoustic frames \cite{sakgoogle}.


In this paper, we present a LSTM based RNN framework which makes more effective
use of model parameters to train prediction models for large scale 
TM prediction. We train and compare our LSTM models at various numbers of parameters and configurations.
We show that LSTM models converge quickly and give state of the
art TM prediction performance for relatively small sized models. Note that we do not address the problem of TM estimation 
in this paper and we suppose that historical TM data is already accurately obtained.

The remainder of this paper is organized as follows: Section \ref{timeseriesprediction} summarizes 
time-series prediction techniques. LSTM RNN architecture and equations are detailed in section \ref{lstm}.
We detail the process of feeding our LSTM architecture and predicting TM in section \ref{tmpredictionusinglstm}.
The prediction evaluation and results are presented in section \ref{experiments}. Related work is presented in section \ref{relatedwork}
and the paper is concluded by section \ref{conclusion}.

\section{Time Series Prediction}\label{timeseriesprediction}

In this section, we give a brief summary of various linear
predictors based on traditional statistical techniques, such as 
ARMA (Autoregressive Moving Average), ARIMA (Autoregressive Integrated Moving Average),
ARAR (Autoregressive Autoregressive) and HW (Holt–Winters) algorithm.
And non-linear time series prediction with neural networks.


\subsubsection{Linear Prediction}


\paragraph{ARMA model}

The time series $\{X_t\}$ is called an ARMA(p, q) process if
$\{X_t\}$ is stationary (i.e. its statistical properties do not change
over time) and 
\begin{equation} \label{delta}
       X_t -  \phi _1 X_{t-1}-. . .- \phi_p X_{t-p} = Z_t +  \theta  _1 Z_{t-1} +. . .+  \theta  _q Z_{t-q}
  \end{equation}

where $\{Z _t\} \approx  WN(0,  \sigma ^2 )$ is white noise with zero mean and
variance $\sigma^2$ and the polynomials  $\phi (z) = 1 -  \phi_1 z - . . . -  \phi_p z^p$
and $\theta (z) = 1 +  \theta_1 z + . . . +  \theta_q z^q$ have no common factors.

The identification of a zero-mean ARMA model which
describes a specific dataset involves the following steps \cite{introductiontimeseries}:
a) order selection (p, q); b) estimation of the mean value of the
series in order to subtract it from the data; c) determination of
the coefficients $\{\phi_i , i = \overline{1,p} \}$ and $\{\theta_i , i =  \overline{1,q}\}$; 
d) estimation of the noise variance $\sigma^2$. 
Predictions can be made recursively using:

$ \widehat{X}_{n+1}= 
 \begin{cases} 
\sum_{j=1}^n   \theta _{nj} (X_{n+1-j}- \widehat{X}_{n+1-j})  &  if  1 \leq n \leq m) \\
\sum_{j=1}^q   \theta _{nj} (X_{n+1-j}- \widehat{X}_{n+1-j})\\
 +  \phi _1 X_n+..+ \phi _p X_{n+1-p} & if  n \geq m
\end{cases} 
$

where $m = max(p, q)$ 
and $\theta_{nj}$ is determined using the
innovations algorithm.
\paragraph{ARIMA model}
A ARIMA(p, q, d) process is described
by:

\begin{equation} \label{phi}
 \phi (B)(1- B)^d X_t = \theta(B)Z_t 
 \end{equation}

where $\phi$ and $\theta$ are polynomials of degree p and q respectively,
$(1 - B)$ represents the differencing operator, d indicates
the level of differencing and B is the backward-shift operator,
i.e. $B^j X_t = X_{t-j} $

\paragraph{ARAR algorithm}
The ARAR algorithm applies memory-shortening transformations,
followed by modeling the dataset as an AR(p)
process: 
$X_t = \phi_1 X_{t-1} + ..+ \phi_p X_{t-p} + Z_t$

The time series $\{Y_t\}$ of long-memory or moderately long-
memory is processed until the transformed series can be
declared to be short-memory and stationary:

\begin{equation} \label{phi}
 S_t = \psi(B)Y_t = Y_t + \psi_1 Y_{t-1} + . . . + \psi_k Y_{t-k}
 \end{equation}
 
The autoregressive model fitted to the mean-corrected series
$X_t = S_t - \overline{S} ̄$, $t =\overline {k + 1, n}$, where $\overline{S}$ 
represents the sample mean for $S_{k+1} , . . . , S_n$ , is given by $\phi(B)X_t = Z_t$ ,
where $\phi (B) = 1- \phi_1 B -  \phi_{l_1} B^{l_1} -  \phi_{l_2} B^{l_2} -  \phi_{l_3} B^{l_3}, \{Z _t\} \approx  WN(0,  \sigma ^2 )$, 
while the coefficients $\phi_j$ and the variance $\sigma^2$ are
calculated using the Yule–Walker equations described in \cite{introductiontimeseries}.
We obtain the relationship:
\begin{equation} \label{phi}
 \xi (B)Y_t =  \phi (1)  \overline{S}  + Z_t 
 \end{equation}
 
where  $\xi (B)Y_t =  \psi (B) \varphi (B) = 1 +  \xi_1B + . . . + \xi_{k+l_3} B^{k+l_3}$
From the following recursion relation we can determine the
linear predictors
\begin{equation} \label{phi}
   P_n Y_{n+h} = - \sum_{j=1}^{k+l_3} \xi P_n Y_{n+h-j} + \phi(1) \overline{S}  \quad h\geq1
 \end{equation}

with the initial condition $P_n Y_{n+h} = Y_{n+h}$ for $h \leq 0$. \\

\paragraph{Holt–Winters algorithm}
The Holt–Winters forecasting algorithm is an exponential
smoothing method that uses recursions to predict the
future value of series containing a trend. 
If the time series has a trend, then the forecast function is:

\begin{equation} \label{phi}
    \widehat{Y}_{n+h} = P_n Y_{n+h} =  \widehat{a}_n +  \widehat{b}_n h 
 \end{equation}

 where $\widehat{a}_n$ and $\widehat{b}_n$ are the estimates of the level of the trend
function and the slope respectively. These are calculated using
the following recursive equations:

\begin{equation} \label{phi}
\begin{cases}
 \widehat{a}_{n+1} =  \alpha Y_{n+1} + (1 -\alpha)( \widehat{a}_n +  \widehat{b}_n) \\
 \widehat{b}_{n+1} =  \beta ( \widehat{a}_{n+1}-  \widehat{a}_n ) + (1 -  \beta )  \widehat{b}_n 
\end{cases} 
\end{equation}

Where $ \widehat{Y}_{n+1} = P_n Y_{n+1} =  \widehat{a}_n +  \widehat{b}_n $ 
represents the one-step
forecast. The initial conditions are:  $\widehat{a}_2 = Y_2$ and 
$\widehat{b}_2 = Y_2 - Y_1$. 
The smoothing parameters $\alpha$ and $\beta$ can be chosen
either randomly (between 0 and 1), or by minimizing the sum
of squared one-step errors  $\sum_{i=3}^n  (Y_i - P_{i-1} Y_i )^2$ \cite{introductiontimeseries}.

\subsubsection{Neural Networks for Time Series Prediction} 

Neural Networks (NN) are widely used for
modeling and predicting network traffic because they can learn
complex non-linear patterns thanks to their strong self-learning and self-
adaptive capabilities. 
NNs are able to estimate almost any linear or non-linear 
function in an efficient and stable manner, when the underlying
data relationships are unknown. 
The NN model is a nonlinear, 
adaptive modeling approach which,
unlike the techniques presented above, relies on the
observed data rather than on an analytical model. 
The architecture and the parameters of the NN are determined
solely by the dataset. 
NNs are characterized by their generalization ability, robustness, fault tolerance,
adaptability, parallel processing ability, etc \cite{studyon}.

\begin{figure}[h] 
\centering
   \includegraphics[scale=0.35]{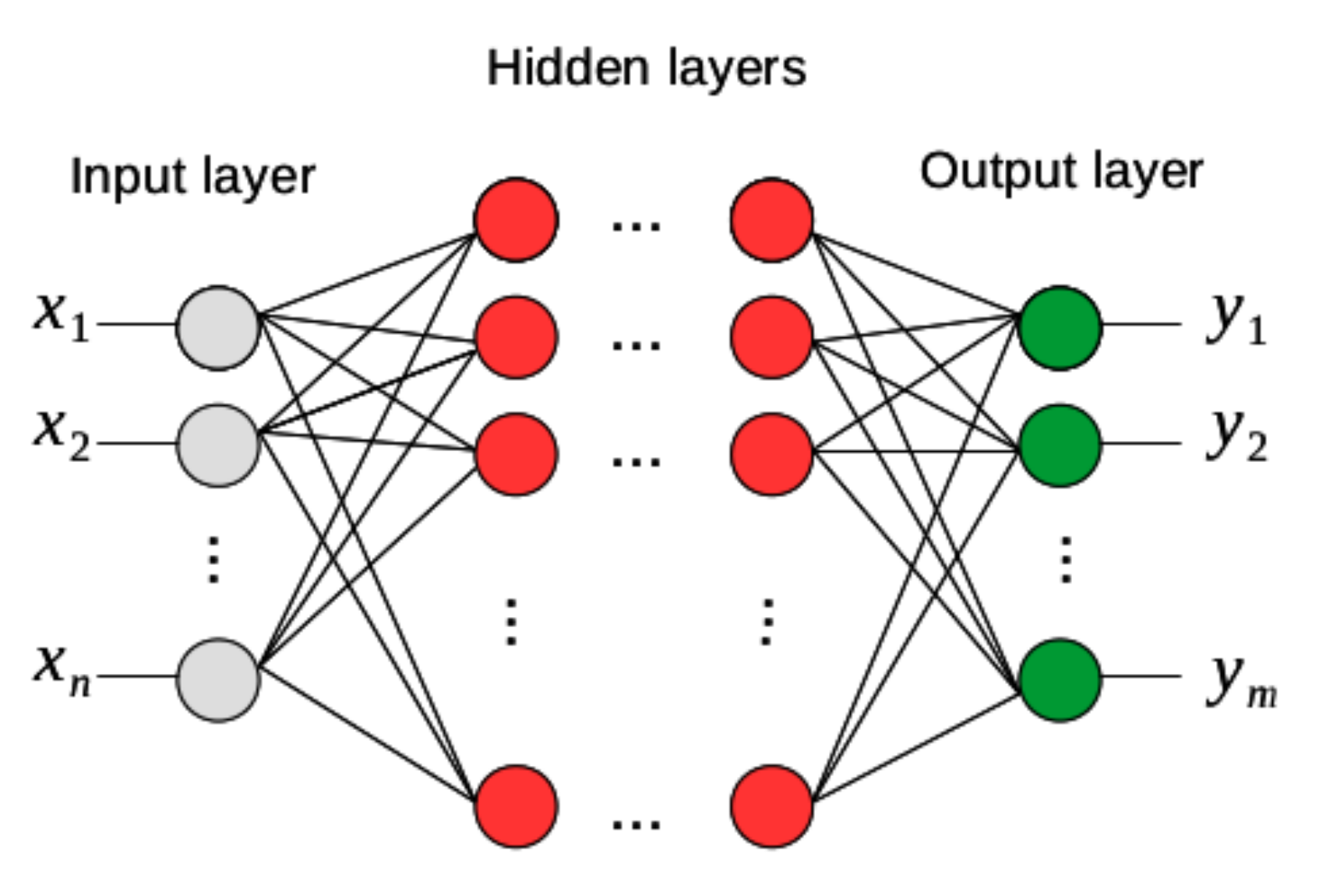}
   \caption{\label{fig:lldppacket} Feed Forward Deep Neural Network}\label{ffnnn}
\end{figure}

A neural network consists of interconnected nodes, called
neurons. The interconnections are weighted and the weights are also called parameters.
Neurons are organized in layers: a) an input layer,
b) one or more hidden layers and c) an output layer. 
The most popular NN architecture is feed-forward in which the
information goes through the network only in the forward
direction, i.e. from the input layer towards the output layer, as
illustrated in figure \ref{ffnnn}.

Prediction using a NN involves two phases: a) the 
training phase and b) the test (prediction) phase. 
During the training phase, the NN is supervised to learn from the data by 
presenting the training data at the input layer and dynamically adjusting 
the parameters of the NN to achieve the desired
output value for the input set. 
The most commonly used
learning algorithm to train NNs is called the backpropagation algorithm. 
The idea of the backpropagation is to propagate of the error backward, 
from the output to the input, where the weights
are changed continuously until the output error falls below a
preset value. 
In this way, the NN learns correlated patterns
between input sets and the corresponding target values. 
The prediction phase represents the testing of the NN. A new unseen input
is presented to the NN and
the output is calculated, thereby predicting the outcome of new
input data.

\section{Long Short Term Memory Neural Networks} \label{lstm}


\begin{figure}[h] 
\centering
   \includegraphics[scale=0.35]{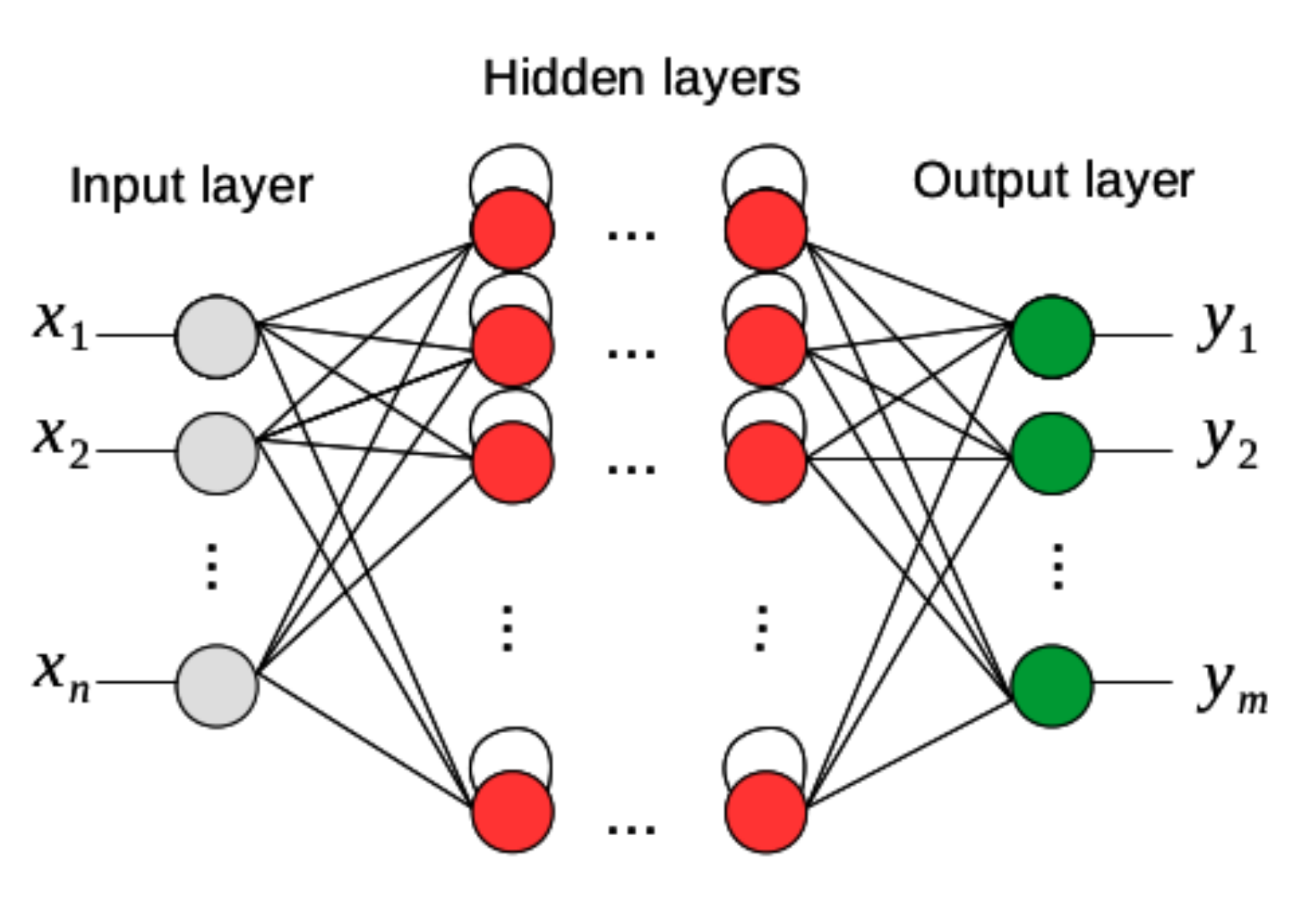}
   \caption{\label{fig:lldppacket} Deep Recurrent Neural Network}\label{drnn}
\end{figure}

\begin{figure}[h] 
\centering
   \includegraphics[scale=0.35]{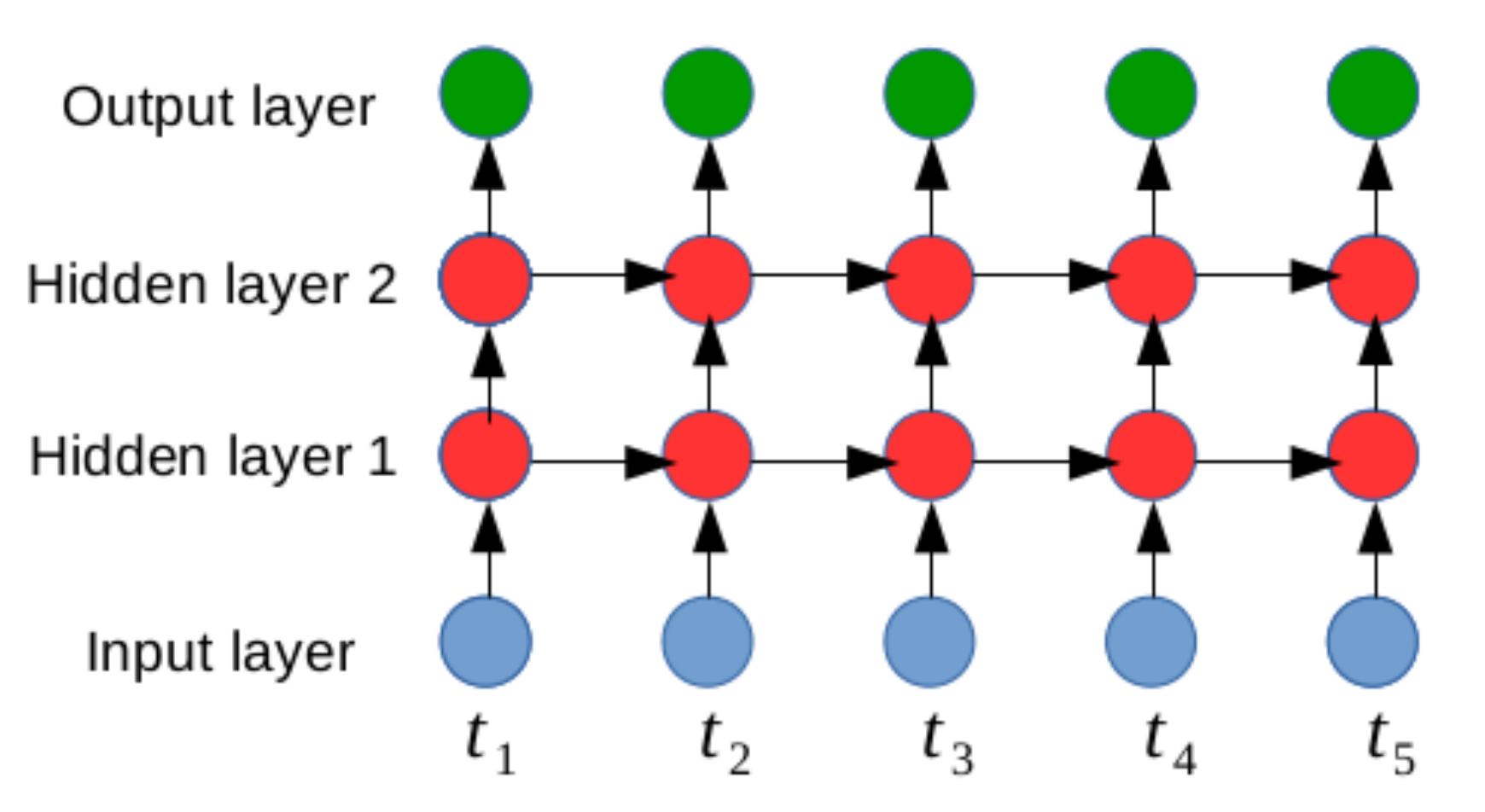}
   \caption{\label{fig:lldppacket} DRNN learning over time}\label{drnnovertime}
\end{figure}

FFNNs can provide only limited temporal modeling by 
operating on a fixed-size window of TM sequence. 
They can only model the data within the window and are unsuited to
handle historical dependencies.
By contrast, recurrent neural networks or 
deep recurrent neural networks (figure \ref{drnn}) contain cycles that feed back
the network activations from a previous time step as inputs to
influence predictions at the current time step (figure \ref{drnnovertime}).
These activations are stored in the internal states of 
the network as temporal contextual information \cite{sakgoogle}.


However, training conventional RNNs with the gradient-based
back-propagation through time (BPTT) technique is difficult due to
the vanishing gradient and exploding gradient problems. 
The  influence  of  a  given  input  on  the  hidden  layers,  and  therefore
on the network output, either decays or blows up exponentially when cycling
around  the  network’s  recurrent  connections. 
These problems limit the capability of RNNs to model the long
range context dependencies to 5-10 discrete time steps between 
relevant input signals and output \cite{sakgooglearxiv}.

To address these problems, an elegant RNN architecture Long Short-Term Memory
(LSTM) – has been designed \cite{lstmoriginalpaper}. 
LSTMs and conventional RNNs have been successfully applied
to sequence prediction and sequence labeling tasks.  LSTM models
have been shown to perform better than RNNs on learning context-
free and context-sensitive languages for example \cite{language}.

\subsection{LSTM Architecture}
\begin{figure}[h] 
\centering
   \includegraphics[scale=0.309]{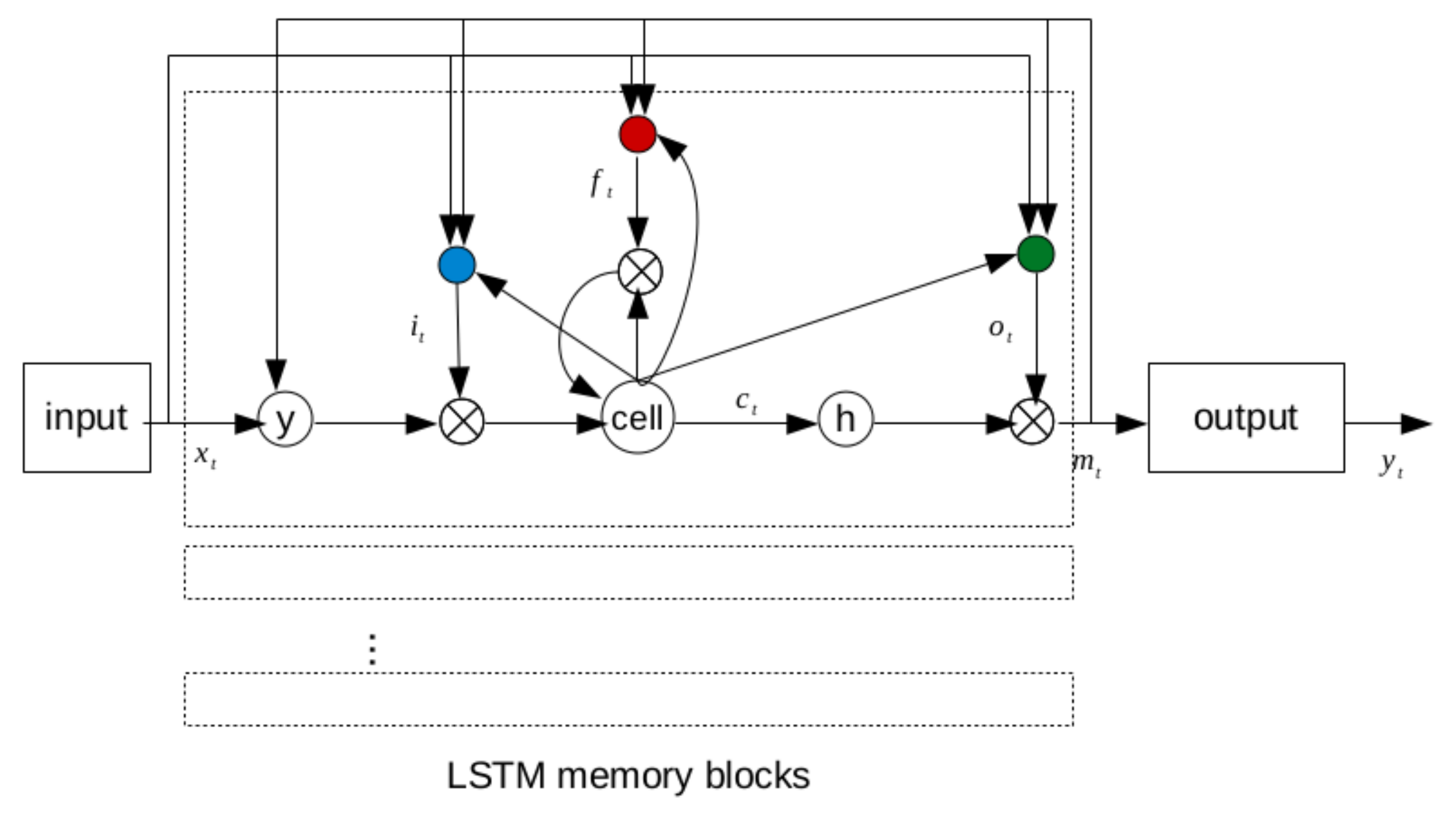}

   \caption{\label{fig:lldppacket} LSTM node} \label{archinode}
\end{figure}

The architecture of LSTMs is composed of units 
called memory blocks.  
Memory block contains memory
cells with self-connections storing (remembering) the temporal state
of the network in addition to special multiplicative units called gates
to control the flow of information.  
Each memory block contains an input gate to control 
the flow of input activations into the memory cell, an 
output gate to control the output flow of cell
activations into the rest of the network and 
a forget gate (figure \ref{archinode}). 

The forget gate scales the internal state of the cell before adding it back to the cell as input through
self recurrent connection, therefore adaptively forgetting
or resetting the cell’s memory. The modern LSTM architecture
also contains peephole connections from its internal cells to the gates in
the same cell to learn precise timing of the outputs \cite{peepholes}.

\subsection{LSTM Equations}


In this subsection we provide the equations for the activation (forward pass) and
gradient calculation (backward pass) of an LSTM hidden layer within a 
recurrent neural network. The backpropagation through time algorithm with
the exact error gradient is used to train the network.
The LSTM equations are given for a single memory block only. 
For multiple blocks the
calculations are simply repeated for each block, in any order \cite{thesisgraves}.

\paragraph{Notations}
\begin{itemize}
 \item \textbf{$w_{ij}$} the weight of the connection from unit i to unit j
 \item \textbf{$a_{i}^t$} the network input to some unit j at time t
 \item \textbf{$b_{i}^t$} the value of the same unit after the activation function has been applied
 \item \textbf{$\iota$} input gate, \textbf{$\phi$} forget gate, \textbf{$\omega $} output gate
 \item \textbf{$C$} set of memory cells of the block
 \item \textbf{$s_c^t$} state of cell $c$ at time $t$ (i.e. the activation of the linear cell unit)
   \item \textbf{$f$} the activation function of the gates, \textbf{$g$} cell input activation functions, \textbf{$h$} cell output activation functions
    \item \textbf{$I$} the number of inputs, \textbf{$K$} the number of outputs, \textbf{$H$} number of cells in the hidden layer
\end{itemize}

Note that only the cell outputs $b_c^t$
are connected to the other blocks in the layer. The other LSTM activations,
such as the states, the cell inputs, or the gate activations, are only visible
within the block. 

We use the index $h$ to refer to cell outputs from other
blocks in the hidden layer.

As with standard RNNs the forward pass is calculated for a length
T input sequence x by starting at t = 1 and recursively applying the update
equations while incrementing t, and the BPTT backward pass is calculated
by starting at t = T , and recursively calculating the unit derivatives while
decrementing t (see Section 3.2 for details). The final weight derivatives are
found by summing over the derivatives at each timestep, as expressed in
Eqn. (3.34). Recall that

\begin{equation} \label{delta}
       \delta_j^t = \frac{\partial O}{\partial a_j^t}
  \end{equation}

Where $O$ is the objective function used for training.
  
  The order in which the equations are calculated during the forward and
backward passes is important, and should proceed as specified below. As
with standard RNNs, all states and activations are set to zero at $t = 0$, and
all $\delta$ terms are zero at $t = T + 1$.

\paragraph{Forward Pass}
\textbf{Input Gates}
\begin{equation} \label{eqa}
     a_\iota^t=\sum_{i=1}^I w_{i\iota} x_i^t+\sum_{h=1}^H w_{h \iota} b_h^{t-1}+\sum_{c=1}^C w_{c \iota} s_c^{t-1}
  \end{equation} 
\begin{equation} \label{eqb}
     b_\iota^t=f(a_\iota^t)
  \end{equation}
  
\textbf{Forget Gates}
\begin{equation} \label{eq1}
     a_\phi^t=\sum_{i=1}^I w_{i\phi} x_i^t+\sum_{h=1}^H w_{h \phi} b_h^{t-1}+\sum_{c=1}^C w_{c \phi} s_c^{t-1}
  \end{equation}
\begin{equation} \label{eqb}
     b_\phi^t=f(a_\phi^t)
  \end{equation}
  
\textbf{Cells}
\begin{equation} \label{eq1}
     a_c^t=\sum_{i=1}^I w_{ic} x_i^t+\sum_{h=1}^H w_{hc} b_h^{t-1}
  \end{equation}
\begin{equation} \label{eq1}
     s_c^t=b_{\phi}^t s_c^{t-1} +b_{\iota}^t g(a_c^t)
  \end{equation}
  
  \textbf{Output Gates}
\begin{equation} \label{eq1}
     a_\omega^t=\sum_{i=1}^I w_{i\omega} x_i^t+\sum_{h=1}^H w_{h\omega} b_h^{t-1}+\sum_{c=1}^C w_{c \omega} s_c^{t-1}
  \end{equation}
\begin{equation} \label{eqb}
     b_\omega^t=f(a_\omega^t)
  \end{equation}
  
\textbf{Cell Outputs}
\begin{equation} \label{eqb}
     b_c^t=b_\omega^t h(s_c^t)
  \end{equation}

\paragraph{Backward Pass}

\begin{equation} \label{delta}
       \epsilon_c^t = \frac{\partial O}{\partial b_c^t}
  \end{equation}
  \begin{equation} \label{delta}
       \epsilon_s^t = \frac{\partial O}{\partial s_c^t}
  \end{equation}

\textbf{Cell Outputs}
\begin{equation} \label{eqa}
     \delta_{\iota}^t = f'(a_{\iota}^t) \sum_{c=1}^C g(a_c^t) \epsilon_s^t
  \end{equation} 
  
 \textbf{Output Gates}
\begin{equation} \label{eq1}
     \delta_{\phi}^t = f'(a_{\phi}^t) \sum_{c=1}^C s_c^{t-1} \epsilon_s^t
  \end{equation}
  
\textbf{States}
\begin{equation} \label{eq1}
     \delta_c^t = b_{\iota}^t g'(a_c^t) \epsilon_s^t
  \end{equation}
  
\textbf{Cells}
\begin{equation} \label{eq1}
     \epsilon_s^t = b_{\omega}^t h'(s_c^t) \epsilon_c^t + b_{\phi}^{t+1} \epsilon_s^{t+1} 
     + w_{c\iota} \delta_{\iota}^{t+1} + w_{c\phi} \delta_{\phi}^{t+1}+  w_{c\omega} \delta_{\omega}^{t+1}
  \end{equation}
  
  \textbf{Forget Gates}
\begin{equation} \label{eq1}
     \delta_{\omega}^t  = f'(a_{\omega}^t) \sum_{c=1}^C h(s_c^t) \epsilon_c^t
  \end{equation}

 \textbf{Input Gates}
\begin{equation} \label{eqb}
     \epsilon_s^t=\sum_{k=1}^K w_{ck} \delta_k^t + \sum_{h=1}^H w_{ch} \delta_h^{t+1}
  \end{equation}

where f(·) (frequently noted as $\sigma(.)$) is the standard logistic sigmoid function defined
in Eq.(8), g(·) and h(·) are the transformations of function
σ(·) whose range are [-2,2] and [-1,1] respectively: 
$\sigma(x)=\frac{1}{1+e^{-x}}$, $g(x)=\frac{4}{1+e^{-x}}-2$ and  $h(x)=\frac{2}{1+e^{-x}}-1$

\section{Traffic Matrix Prediction Using LSTM RNN} \label{tmpredictionusinglstm} 

In this section we describe the use of a deep LSTM architecture with a 
deep learning method to extract the dynamic features
of network traffic and predict the future TM. This architecture can deeply excavate mutual
dependence among the traffic entries in various timeslots.

\subsection{Problem Statement} \label{problemstatement}
Let N be the number of nodes in the network. The $N$-by-$N$ traffic matrix is denoted by Y 
such as an entry $y_{ij}$ represents the traffic volume flowing from node i to node j. 
We add the time dimension to obtain a structure of N-by-N-by-T tensor (vector of matrices) S such as an entry 
$s_{ij}^t$ represents the volume of traffic flowing from node i to node j at time t,
and T is the total number of time-slots.
The traffic matrix prediction problem is
defined as solving the predictor of $Y^t$ (denoted by $\widehat{Y}^t$) via a series of
historical and measured traffic data set ($Y^{t-1}$, $Y^{t-2}$, $Y^{t-3}$, ..., $Y^{t-T}$).
The main challenge here is how to model the inherent 
relationships among the traffic data set so that one can exactly predict $Y^{t}$.

\subsection{Feeding The LSTM RNN} \label{feeding}

To effectively feed the LSTM RNN, we transform each matrix $Y^t$  to a vector $X^t$ (of size $N\times N$) 
by concatenating its N rows from top to bottom. $X^t$ is called traffic vector (TV).
Note that $x_n$ entries can be mapped to the original $y_{ij}$ using the relation $n=i\times N+j$.
Now the traffic matrix prediction problem is
defined as solving the predictor of $X^t$ (denoted by $\widehat{X}^t$) via a series of
historical measured traffic vectors ($X^{t-1}$, $X^{t-2}$, $X^{t-3}$, ..., $X^{t-T}$).

One possible way to predict the traffic vector $X^t$ is to predict one component $x_n^t$ at a time 
by feeding the LSTM RNN one vector ($x_0^t, x_1^t, ..., x_{N^2}^t)$ at a time.
This is based on the assumption that each OD traffic is independent from all other ODs 
which was shown to be wrong by \cite{nongaussian}. Hence, considering the previous traffic of all 
ODs is necessary to obtain a more accurate prediction of the traffic vector.

\textbf{Continuous Prediction Over Time}:
Real-time prediction of traffic matrix requires continuous feeding and learning. 
Over time, the total number of time-slots become too big resulting in high computational complexity.
To cope with this problem, we introduce the notion of learning window (denoted by $W$) which indicates a fixed number of previous
time-slots to learn from in order to predict the current traffic vector $X^t$ (Fig. \ref{window}).

\begin{figure}[h] \label{window}
\centering
   \includegraphics[scale=0.36]{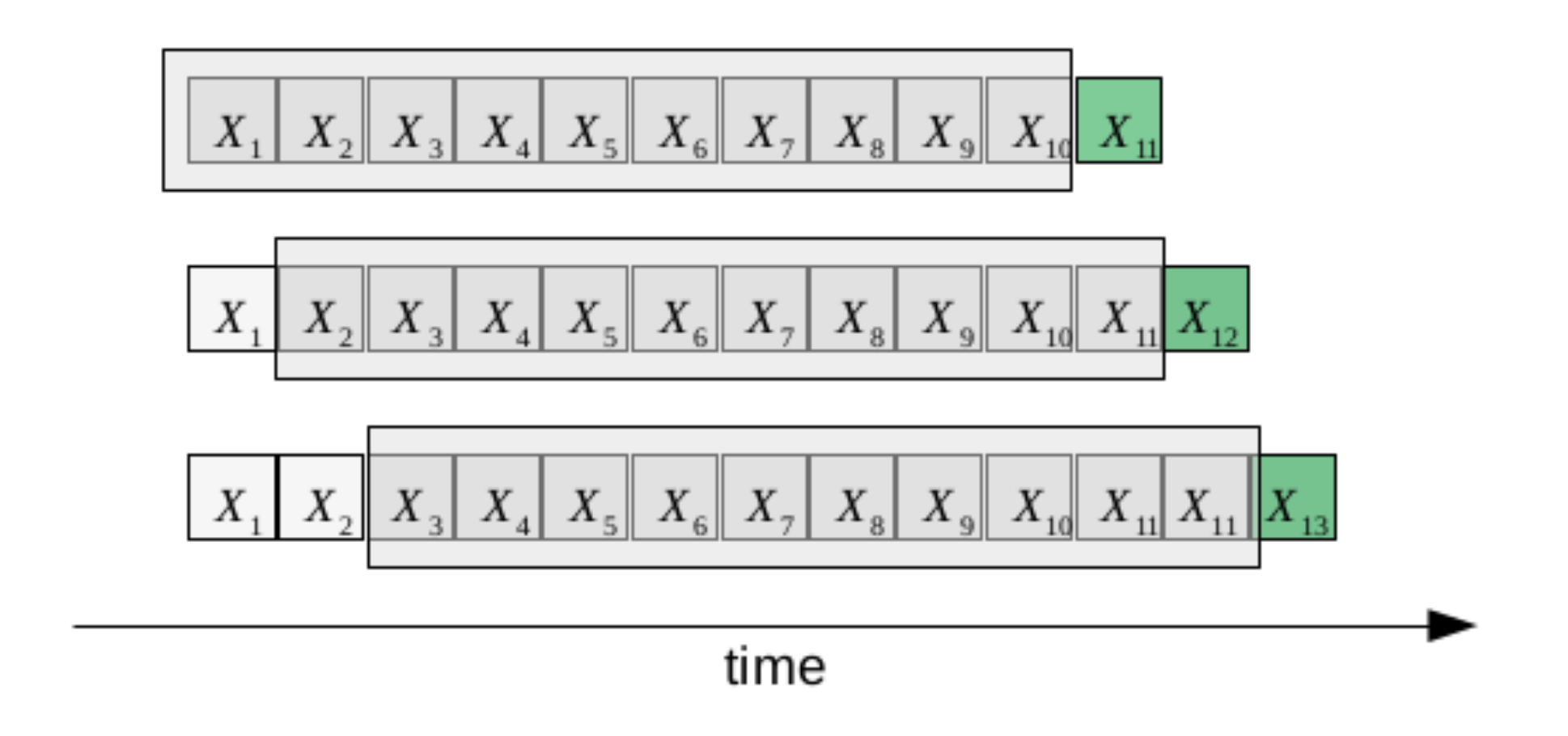}
   \caption{\label{fig:lldppacket} Sliding learning window}\label{window}
\end{figure} 

We construct the $W$-by-$N^2$ traffic-over-time matrix (that we denote by $M$) by putting together $W$ vectors 
($X^{t-1}$, $X^{t-2}$, $X^{t-3}$, ..., $X^{t-W}$) ordered by time. 
Note that $T\geq W$ ($T$ being the total number of historical matrices) and the number of matrices $M$ is equal to $T/W$.

\subsection{Performance Metric}
To quantitatively assess the overall performance of our LSTM model, 
Mean Square Error (MSE) is used to estimate the prediction accuracy.
MSE is a scale dependent metric
which quantifies the difference between the forecasted
values and the actual values of the quantity being 
predicted by computing the average sum of squared errors:

\begin{equation} \label{phi}
MSE=\frac{1}{N} \sum_{i=1}^N (y_i- \widehat{y}_i )^2
\end{equation}
where $y_i$ is the observed value, $\widehat{y}_i$ is the predicted value
and N represents the total number of predictions.

\section{Experiments and Evaluation} \label{experiments}

In this section, we will evaluate the prediction accuracy 
of our method using real traffic data from the G\'EANT 
backbone networks \cite{geant}. 
G\'EANT is the pan-European research network. 
G\'EANT has a PoP in each European
country and it carries research traffic from the European National
Research and Education Networks (NRENs) connecting universities and research institutions. 
As of 2005, the G\'EANT network was made up of 23 peer nodes 
interconnected using 38 links. 
In addition,
G\'EANT has 53 links with other domains.

 
2004-timeslot traffic matrix data is sampled from the G\'EANT network by 15-min
interval \cite{geantdata} for several months. 
In our simulation, we also compare our prediction and
estimation methods with a state-of-the-art method, that is, the PCA
method introduced in the above section.

To evaluate our method on short term traffic matrix prediction, 
we consider a set of 309 traffic
matrices measured between 01-01-2005 00am and 04-01-2005 5:15am.
As detailed in section \ref{feeding}, we transform the matrices to vectors of size $529$ each and we concatenate 
the vectors to obtain the traffic-over-time matrix $M$ of size $309\times 529$. We split $M$ into
two matrices, training matrix $M_{train}$ and validation matrix $M_{test}$ of sizes $263$ and $46$ consecutively. 
$M_{train}$ is used to train the LSTM RNN model and $M_{test}$ is used to evaluate and validate its accuracy. Finally, 
We normalize the data by dividing by the maximum value. 

\begin{figure}[h] 
\centering
   \includegraphics[scale=1.4]{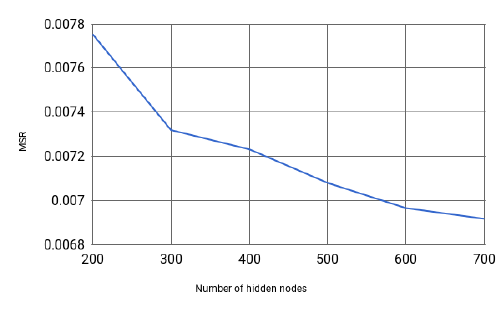}
   \caption{\label{fig:lldppacket} MSE over size of hidden layer}\label{mstbynhidden}
    \vspace{-1em}
\end{figure}

\begin{figure}[h] 
\centering
   \includegraphics[scale=1.4]{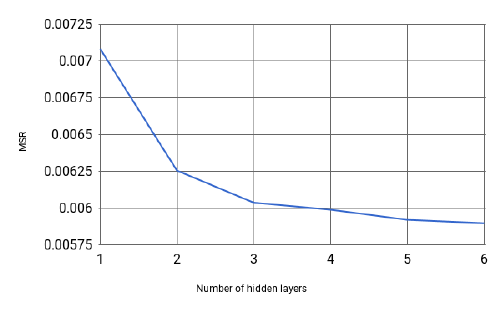}
   \caption{\label{fig:lldppacket} MSE over number of hidden layers (500 nodes each)}\label{msrbynhiddenlayers}
  \vspace{-1em}
\end{figure}

\begin{figure}[h] 
\centering
   \includegraphics[scale=1.4]{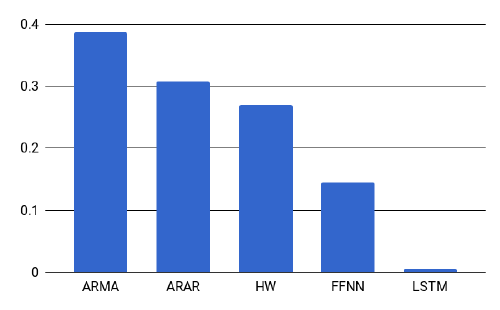}
   \caption{\label{fig:lldppacket} Comparison of prediction methods}\label{compare}
\end{figure}

We use Keras library \cite{keras} to build and train our model. 
The training is done on a Intel core i7 machine with 16GB memory. 
Figures \ref{mstbynhidden} and \ref{msrbynhiddenlayers} show the variation of the prediction error when using different 
numbers of hidden units and hidden layers respectively. Finally, figure \ref{compare} compares the prediction error of 
the different prediction methods presented in this paper and shows the superiority of LSTM. Note that, the prediction 
results of the linear predictors and FFNN are obtained from \cite{evaluation} and they represent the error of predicting 
only one traffic value which is obviously an easier task than predicting the whole traffic matrix.

\section{Related Work} \label{relatedwork}
Various methods have been proposed to predict traffic matrix. 
\cite{evaluation} evaluates and compares traditional linear prediction models (ARMA, ARAR, HW)
and neural network based prediction with multi-resolution learning. 
The results show that NNs outperform
traditional linear prediction methods
which cannot meet the accuracy requirements.
\cite{nongaussian} proposes a FARIMA predictor 
based on an $\alpha$-stable non-Gaussian self-similar traffic model. 
\cite{predictionandcorrection} compares three prediction methods: 
Independent Node Prediction (INP), 
Total Matrix Prediction with Key Element Correction (TMP-KEC)
and Principle Component Prediction with Fluctuation Component Correction (PCP-FCC). 
INP method does
not consider the correlations among the nodes, resulting in
unsatisfying prediction error. TMP-KEC method reduces the
forecasting error of key elements as well as that of the total
matrix. PCP-FCC method improves the overall prediction error
for most of the OD flows.

\section{Conclusion}\label{conclusion}

In this work, we have shown that LSTM RNN architectures are well suited for traffic matrix prediction.
We have proposed a data pre-processing and RNN feeding technique that achieves high prediction accuracy 
in a few seconds of computation (approximately 60 seconds for one hidden layer of 300 nodes). The results of our evaluations show that LSTM RNNs outperforms
traditional linear methods and feed forward neural networks by many orders of magnitude.

\end{document}